\documentclass[aps,preprint,prd,showpacs,nofootinbib]{revtex4}
\usepackage{latexsym}
\usepackage{amsmath}
\usepackage{graphicx}
\usepackage{subfigure}
\usepackage{dcolumn}
\usepackage{bm}
\usepackage{amssymb}
\usepackage{latexsym}
\usepackage{ulem}
\usepackage{color}
\usepackage{afterpage}
\usepackage[colorlinks,linkcolor=magenta,anchorcolor=cyan,citecolor=blue]{hyperref}

\def\be{\begin{equation}}
\def\ee{\end{equation}}
\def\ba{\begin{eqnarray}}
\def\ea{\end{eqnarray}}

\def\nn{\nonumber}
\def\lf{\left}
\def\rt{\right}

\begin{document}

\title{Propagating speed of primordial gravitational waves and inflation }

\author{Yong Cai$^{1}$\footnote{caiyong13@mails.ucas.ac.cn}}
\author{Yu-Tong Wang$^{1}$\footnote{wangyutong12@mails.ucas.ac.cn}}
\author{Yun-Song Piao$^{1}$\footnote{yspiao@ucas.ac.cn}}

\affiliation{$^1$ School of Physics, University of Chinese Academy of
Sciences, Beijing 100049, China}

%\affiliation{$^2$ State Key Laboratory of Theoretical Physics, Institute of Theoretical Physics, \\
%Chinese Academy of Sciences, P.O. Box 2735, Beijing 100190, China}

\begin{abstract}

We show that if the propagating speed of gravitational waves (GWs)
gradually diminishes during inflation, the power spectrum of
primordial GWs will be strongly blue, while that of the primordial scalar
perturbation may be unaffected.  We also illustrate that such a
scenario is actually a disformal dual to the superinflation, but
it does not have the ghost instability. The blue tilt obtained is
$0<n_T\lesssim 1$, which may significantly boost the stochastic
GWs background at the frequency band of Advanced LIGO/Virgo, as
well as the space-based detectors.

\end{abstract}

\maketitle

\section{Introduction}

Recently, the LIGO Scientific Collaboration has observed a
transient gravitational wave (GWs) signal with a significance in
excess of 5.1$\sigma$ \cite{Abbott:2016blz}, which is consistent
with an event of the binary black hole coalescence. This
discovery will be a scientific milestone for understanding our
universe, if it is confirmed.

It is speculated that the stochastic GWs background contributed by
the incoherent superposition of all merging binaries in the
universe might be higher than expected previously
\cite{TheLIGOScientific:2016wyq}, which is potentially measurable
around $25$Hz by the Advanced LIGO/Virgo detectors operating at
their projected final sensitivity.
%Though LIGO's GWs signal could be attributed to astrophysical
%sources, as has been inferred,
However, some cosmological sources may also contribute a
stochastic background of GWs at the corresponding frequency band, such
as cosmic strings \cite{Damour:2000wa} and cosmological phase
transitions \cite{Kamionkowski:1993fg}\cite{Dev:2016feu}.

%Inspired by the LIGO's discovery, we would like to recheck if the
%stochastic background of primordial GWs from inflation is actually
%negligible.

It is well known that the standard slow-roll inflation predicts a
nearly flat spectrum of scalar perturbation, as well as primordial
GWs \cite{Starobinsky:1979ty}\cite{Rubakov:1982}. Recently, the
BICEP2/Keck data, combined with the Planck data and the WMAP data,
have put the constraint $r<0.09$ (95\% C.L.) \cite{Array:2015xqh}
on the amplitude of primordial GWs on large scale, or at ultra-low
frequency, which corresponds to $\Omega_{gw}\sim 10^{-15}$, but there is no
strong limit for its tilt $n_T$. Actually, as long as its spectrum
is  blue enough, the stochastic GWs background from primordial
inflation is also not negligible at the frequency band of Advanced
LIGO/Virgo.

The slow-roll inflation model with $\epsilon = -{\dot H}/H^2\ll 1$
generally has $n_T=-2\epsilon < 0$. Thus $n_T>0$ requires either
the superinflation \cite{Piao:2004tq}\cite{Baldi:2005gk}, also
\cite{Liu:2013iha}\cite{Cai:2015yza}, which breaks the null energy condition (NEC), or an
anisotropic stress source during inflation, e.g., the particle
production
\cite{Cook:2011sp}\cite{Sorbo:2011ja}\cite{Mukohyama:2014gba}\cite{Namba:2015gja}.
During the superinflation, the primordial GWs come from the
amplification of vacuum tensor perturbations. However, since the
almost scale-invariance of the scalar perturbation requires
$|\epsilon|\sim 0.01 $, we generally have $|n_T|\sim {\cal
O}(0.01)$ for the superinflation. Obtaining a blue GWs
spectrum $n_T>0.1$ without the ghost instability while reserving a
scale-invariant scalar spectrum with slightly red tilt is still a
challenge for the inflation scenario\footnote{It is found that in the pre-big bang scenario (obtained in the context of string cosmology) the primordial GWs spectrum is blue \cite{Brustein:1995ah}\cite{Gasperini:2002bn}\cite{Gasperini:2007zz}.}, see e.g.\cite{Wang:2014kqa}
for comments.

In Einstein gravity, the propagating speed $c_T$ of GWs is the same as
the speed of light, thus can naturally be set as unity.
Nevertheless, it might be modified when dealing with the extremely
early universe, e.g., the low-energy effective string theory with
higher-order corrections
\cite{Met:1987}\cite{Antoniadis:1993jc}\cite{Kawai:1998ab}\cite{Cartier:1999vk},
see also \cite{Maeda:2004vm}\cite{Nojiri:2015qyc}. Since the
amplitude of the primordial GWs is determined by $c_T$ and the
Hubble radius $\sim H^{-1}$, the running of $c_T$ will inevitably
affect the power spectrum of primordial GWs (see also \cite{Giovannini:2015kfa} for the study from the point of view of the running of GWs' refractive index $n$). It was found in
\cite{Cai:2015ipa}\cite{Cai:2015dta} that the oscillation of $c_T$
may leave some observable imprints in CMB B-mode polarization. The
effect of the sound speed $c_S$ of scalar perturbation on the
scalar spectrum has been investigated in e.g.\cite{Khoury:2008wj}
\cite{Park:2012rh}.

Here, we show that if the propagating speed $c_T$ of GWs gradually
diminishes during inflation, the power spectrum of primordial GWs
will be strongly blue, while the spectrum of scalar perturbation may
be still that of slow-roll inflation. There is no the ghost
instability. The blue tilt obtained is $0<n_T\lesssim 1$, which
may significantly boost the stochastic GWs background within the
window of Advanced LIGO, as well as those of the space-based detectors.

%Here, we propose a design for inflation, in which $c_T$ is
%gradually diminishing. We find that the GWs spectrum may be
%blue-tilt with $n_T>0.1$, while the scalar spectrum still
%maintains the slightly red tilt of slow-roll inflation. Thus our
%model, consistent with the observation at CMB scale, has the
%signal of primordial GWs at recent LIGO's regime.

%The propagating speed of primordial GWs encodes the information of
%modified gravity during inflation, our work

%suggests that provide valuable insights to the gravity physics at
%extremely high energy scale.

\section{Inflation and $c_T$}

\subsection{The model}\label{model}

We follow the effective field theory of inflation
\cite{Cheung:2007st}, beginning  with the Langrangian in unitary gauge
\ba S & = & {M_p^2\over 2}\int d^4x \sqrt{-g} \Big[R-c_1(t)-c_2(t)g^{00}\label{action1}\\
& &\,\,\,\,\,\,\,\,\,\,\,\,\,\,\,\,\,\,\,\,\,\,\,\,\,\,\,\,\,- \,
\left(1-{1\over c_T^{2}(t)}\right)\left(\delta K_{\mu\nu}\delta
K^{\mu\nu}-\delta K^2 \right)  \Big]\,, \label{action} \ea
where $M_p=1/\sqrt{8\pi G}$,
$c_1(t)=2({\dot H}+3H^2)$, $c_2(t)=-2{\dot H}$, and a dot denotes the derivative with respect to cosmic time $t$. We will work in
the inflation background with $0<\epsilon\ll 1$, which may be set
by requiring $|{\dot H}|\ll H^2$ in (\ref{action1}). The scalar
perturbation at quadratic order is not affected by $\delta
K_{\mu\nu}\delta K^{\mu\nu}-\delta K^2$, see Appendix
\ref{scalar}, and also \cite{Creminelli:2014wna}, so its spectrum
is determined by slow-roll parameters. However, the quadratic
action of tensor perturbation is altered as\footnote{In \cite{Giovannini:2015kfa}, the author investigated the effect induced by the running of GWs' refractive index $n(\tau)$, which is similar to that of $c_T$. But note that $c_T\neq 1/n$, as can be seen from the difference between Eq.(\ref{paction}) here and the Eq.(2.8) in \cite{Giovannini:2015kfa}.}
\be
S^{(2)}_{\gamma}=\int d\tau d^3x {M_p^2a^2c_T^{-2}\over
8}\lf[\lf(\frac{d\gamma_{ij}
}{d\tau}\rt)^2-c_T^2(\vec{\nabla}\gamma_{ij})^2\rt]\,,
\label{paction}\ee
where $\tau=\int dt/a$, and $\gamma_{ij}$
satisfies $\gamma_{ii}=0$ and $\partial_i \gamma_{ij}=0$.

The Fourier series of $\gamma_{ij}$ is \be
\gamma_{ij}(\tau,\mathbf{x})=\int \frac{d^3k}{(2\pi)^{3}
}e^{-i\mathbf{k}\cdot \mathbf{x}} \sum_{\lambda=+,\times}
\hat{\gamma}_{\lambda}(\tau,\mathbf{k})
\epsilon^{(\lambda)}_{ij}(\mathbf{k}), \ee in which $
\hat{\gamma}_{\lambda}(\tau,\mathbf{k})=
\gamma_{\lambda}(\tau,k)a_{\lambda}(\mathbf{k})
+\gamma_{\lambda}^*(\tau,-k)a_{\lambda}^{\dag}(-\mathbf{k})$, the
polarization tensors $\epsilon_{ij}^{(\lambda)}(\mathbf{k})$
satisfy $k_{j}\epsilon_{ij}^{(\lambda)}(\mathbf{k})=0$,
$\epsilon_{ii}^{(\lambda)}(\mathbf{k})=0$,
 and $\epsilon_{ij}^{(\lambda)}(\mathbf{k})
\epsilon_{ij}^{*(\lambda^{\prime}) }(\mathbf{k})=\delta_{\lambda
\lambda^{\prime} }$, $\epsilon_{ij}^{*(\lambda)
}(\mathbf{k})=\epsilon_{ij}^{(\lambda) }(-\mathbf{k})$, the
annihilation and creation operators $a_{\lambda}(\mathbf{k})$ and
$a^{\dag}_{\lambda}(\mathbf{k}^{\prime})$ satisfy $[
a_{\lambda}(\mathbf{k}),a_{\lambda^{\prime}}^{\dag}(\mathbf{k}^{\prime})
]=\delta_{\lambda\lambda^{\prime}}\delta^{(3)}(\mathbf{k}-\mathbf{k}^{\prime})$.
The equation of motion for $u(\tau,k)$ is
\be
\frac{d^2u}{d\tau^2}+\left(c_T^2k^2-\frac{d^2z_T/d\tau^2}{z_T}
\right)u=0, \label{eom1}
\ee
where \be {u}(\tau,k)=
\gamma_{\lambda}(\tau,k) {z_T}, \quad z_T= {aM_p { c_T^{-1} }\over
2}.\label{zt} \ee Initially, the perturbations are deep inside the
sound horizon, i.e., $c_T^2k^2 \gg \frac{d^2z_T/d\tau^2}{z_T}$,
the initial state is the Bunch-Davies vacuum, thus $u\sim
\frac{1}{\sqrt{2c_T k} }e^{-i c_T k\tau}$. The power spectrum of
primordial GWs is \be
P_T=\frac{k^3}{2\pi^2}\sum_{\lambda=+,\times} \lf|\gamma_{\lambda}
\rt|^2=\frac{4k^3}{\pi^2 M_p^2}\cdot\frac{c_T^2}{ a^2} \lf|u
\rt|^2, \quad aH/(c_Tk) \gg 1.\label{pt} \ee

The diminishment of $c_T$ may be regarded as \be c_T=
(-H_{inf}\tau)^p, \label{cTn}\ee in which $p>0$, and $H_{inf}$ is
the Hubble parameter during inflation, which is regarded as
constant for simplicity. Additionally, Eq. (\ref{cTn}) suggests ${{\dot c}_T\over
H_{inf}c_T}=-p$.

We set $dy=c_Td\tau$, thus Eq. (\ref{eom1}) is rewritten as
\be
u_{,yy}+\left(k^2-\frac{z_{T,yy}}{z_T} \right)u=0, \label{eom4} \ee
where ${u}(y,k)= \gamma_{\lambda}(y,k) {z_T}$, $z_T= {aM_p {
c_T^{-1/2} }\over 2}$ and the subscript `$,y$' denotes $d/dy$. Note here $u(y,k)$ and $z_T$ are different from those in Eq. (\ref{eom1}), but $\gamma_{\lambda}$ is still the same.
The solution of Eq.(\ref{eom4}) is \be
u_k(y)={\sqrt{\pi}\over2\sqrt{k}}\sqrt{-ky}H^{(1)}_{\nu}(-ky),\label{cq002}
\ee
%where $\nu={p+3/2\over p+1}$ and
%\be
%H^{(1)}_{\nu}(-ky)\overset{-ky\rightarrow0}\approx-i\lf({2\over
%-ky}\rt)^{1+{1\over 2(1+p)}}{\Gamma\lf(1+{1\over 2(1+p)}\rt)\over
%\pi}\,\,.
%\ee
where \be
H^{(1)}_{\nu}(-ky)\overset{-ky\rightarrow0}\approx-i\lf({2\over
-ky}\rt)^{\nu}{\Gamma\lf(\nu\rt)\over \pi}\,\,, \ee and
$\nu=1+{1\over 2(1+p)}$.  Thus the spectrum (\ref{pt}) is
\be
P_T={4k^3\over \pi^2 M_P^2} {c_T|u|^2\over a^2}=
{2^{-p\over1+p}\over \pi }\Gamma^2 \lf({1\over2(1+p)}\rt)
{2H_{inf}^2\over \pi^2 M_P^2 c_T} (-ky)^{p\over 1+p},
\label{nT}
\ee
where $y={c_T\tau\over {1+p} }={-}{c_T\over
{(1+p)}aH_{inf} }$. Therefore,
\be n_T={p\over 1+p} \label{nTB}\ee is
blue-tilt, which is $n_T\simeq p$ for $p\ll 1$ and $n_T\simeq 1$
for $p\gg 1$. Here, the running of $H_{inf}$ may contribute
$-2\epsilon\sim -0.01$, which has been neglected.

Thus, we obtain a blue-tilt GWs spectrum with $0<n_T\leqslant 1$.
Here, both the scalar perturbation and the background are unaffected by additional operator (\ref{action}). The background is
set by (\ref{action1}), which is the slow-roll inflation with
$0<\epsilon\ll 1$, so the scalar spectrum is flat with a slightly
red tilt, which is consistent with the observations. It is
noticed that based on the effective field theory of inflation, the
introducing of other operators may also result in the blue-tilt
GWs spectrum \cite{Cannone:2014uqa}\cite{Baumann:2015xxa},
however, in \cite{Cannone:2014uqa} $n_T>0.1$ requires that the
graviton have a large mass $m_{graviton}\simeq H_{inf}$, while in
\cite{Baumann:2015xxa} $|{{\dot c}_T\over H_{inf}c_T}|\ll 1$ was
implicitly assumed.

It is well known that the blue-tilt GWs spectrum is the hallmark
of the superinflation. Here, the scenario proposed is actually a
disformal dual to the superinflation.
We will discuss this issue
in detail in Sec. \ref{app-disformal}.

%When $p\ll 1$, i.e., $c_T$ is slowly varying, the result is
%consistent with that in Ref.\cite{DeFelice:2014bma}. However, ours
%is valid not only for $p\ll 1$, but also for arbitrary positive
%value of $p$. In addition, it should be mentioned that
%$0<n_T\leqslant 1$, i.e., $n_T$ is not able to be larger than 1.

\subsection{The stochastic background of GWs}\label{twoB}

We will focus on the stochastic background of GWs from such a
scenario of inflation. The present observations are still not able to put stringent constraints on $c_T$ at present (see, e.g., \cite{Amendola:2014wma}\cite{Raveri:2014eea}, also \cite{Moore:2001bv} for the constraint on the phase velocity and \cite{Blas:2016qmn} for the group velocity).
Future observations may put more stringent constraints on $c_T$ \cite{Nishizawa:2014zna}\cite{Nishizawa:2016kba}.
However, we will not get involved in this issue too much and we will assume that $c_T(t)$ will return to $c_T=1$ at certain time before
the end of inflation.
Conventionally, one define
\begin{equation}
\label{density} \Omega_{\text{gw}}(k,
\tau_{0})=\frac{1}{\rho_{\text{c}}}\frac{d\rho_{\text{gw}}}{d\ln
k}=\frac{k^{2}}{12 a^2_0H^2_0}P_{T}T^2(k,\tau_0)\,,
\end{equation}
where $\rho_{\text{c}}=3H^{2}_0/\big(8\pi G\big)$, $\tau_{0}=1.41\times10^{4}$ Mpc, $a_0=1$, $H_0=67.8$ km s$^{-1}$
Mpc$^{-1}$,  the reduced Hubble parameter $h=H/\big(100\,
\text{km s}^{-1}\text{Mpc}^{-1}\big)$, and $\rho_{\text{gw}}$ is
the energy density of relic GWs at present, so $\Omega_{gw}(k,
\tau_{0})$ reflects the fraction of $\rho_{\text{gw}}$ per
logarithmic frequency interval. The transfer function is
\cite{Turner:1993vb}\cite{Boyle:2005se}\cite{Zhang:2006mja}
\be
T(k,\tau_{0})=\frac{3
\Omega_{\text{m}}j_1(k\tau_0)}{k\tau_{0}}\sqrt{1.0+1.36\frac{k}{k_{\text{eq}}}+2.50\big(\frac{k}{k_{\text{eq}}})^{2}},
\label{Tk}
\ee
where $k_{\text{eq}}=0.073\,\Omega_{\text{m}} h^{2}$
Mpc$^{-1}$ is that of the perturbation mode that entered the
horizon at the equality of matter and radiation. We have neglected
the effects of the neutrino free-streaming on $T(k,\tau_0)$
\cite{Weinberg:2003ur}, which is actually negligible.
The underlying assumption on the thermal history of the post-inflation
universe is able to affect $T(k,\tau_{0})$  significantly, see
e.g.\cite{Kuroyanagi:2014nba}, but we will only focus on the
simplest case described by Eq.(\ref{Tk}).

One generally parameterizes $P_T$ as
\be  P_T =
A_T\left(\frac{k}{k_*}\right)^{n_T}, \label{para1}
\ee
where
$k_*=0.01$ Mpc$^{-1}$ is the pivot scale. However, if
$n_T>0.4$, one will have $P_T>1$ at high-frequency region
($f>10^5$Hz). The GWs with $P_T\sim 1$ will induce the same-order
scalar perturbation at nonlinear order, e.g.\cite{Wang:2014kqa},
which will result in the overproduction of primordial black holes at
the corresponding scale, which is inconsistent with their
abundance. The upper bound put by the production of primordial
black holes is $P_T<0.4$ \cite{Nakama:2015nea}. In addition, the
indirect upper bound given by the combination of CMB with lensing,
BAO and BBN observations is $\Omega_{gw}<3.8\times 10^{-6}$
\cite{Pagano:2015hma}, which also puts a strong constraint on $n_T$,
i.e., $n_T<0.36$ at $95\%$ C.L. for $r=0.11$ \cite{Lasky:2015lej},
otherwise $\Omega_{gw}$ at higher frequency will exceed this
bound.

However, in our scenario, $c_T(t)$ is assumed to return to unity at a certain
time $t_c$ before the end of inflation, as has
been mentioned. This means that the blue-tilt spectrum will
acquire a cutoff around $k_c$, see Sec. \ref{app-cutoff} for details,
which may avoid the above constraints on $n_T$. We may
parameterize the corresponding $P_T$ as
\ba P_{T} =
A_T\left[1-e^{-\left(\frac{k}{k_c}\right)^{n_T}}\right]
\left(\frac{k_c}{k_*}\right)^{n_T}, \label{para2}
\ea
which is
(\ref{para1}) for $k\ll k_c$, and tends to a constant
$A_T(\frac{k_c}{k_*})^{n_T}$ for $k\gg k_c$. Though we will use
(\ref{para1}) and (\ref{para2}) since we are mainly interested in the boosted blue-tilted spectrum,  we should  point out that $P_T$ will decrease at $k>k_c$ or $k\gg k_c$ (which may be out of the range we are interested in),  if we assume that $c_T$ will increase back to unity.
In such case, $P_T$ may be parameterized as
\be P_T= A_T
\left(\frac{k}{k_*}\right)^{n_T} { 1\over
1+\left({k}/{k_c}\right)^{n_{Tc}}}, \label{para3}
\ee
where
$n_{Tc}> n_T$, so that when $k\gg k_c$, $P_T=A_T
({k_c}/{k_*})^{n_T}({k}/{k_c})^{n_T-n_{Tc}}$ has a red tilt. When
$n_{Tc}=n_T$, (\ref{para3}) is similar to (\ref{para2}).

%The model with $k_c$ is discussed in detail in Appendix B.

\begin{figure}[htbp]
\includegraphics[scale=2,width=0.55\textwidth]{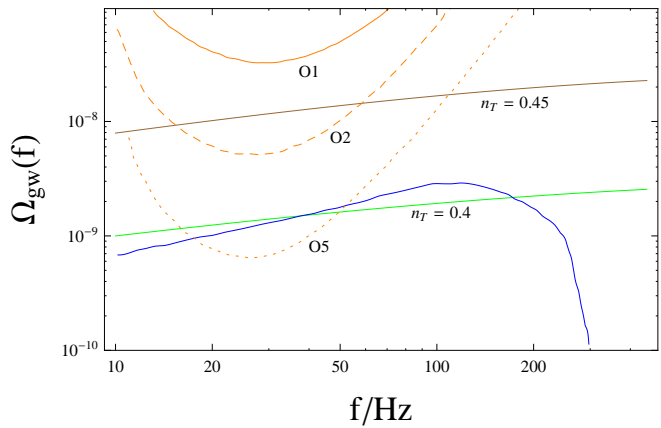}
\caption{The brown line is the stochastic GWs background from
inflation with spectral index $n_T=0.45$ and tensor-to-scalar ratio $r=0.05$ at the CMB scale. O1, O2, and O5 curves, taken from
\cite{TheLIGOScientific:2016wyq}, are the current Advanced
LIGO/Virgo sensitivity, the observing run (2016-2017) and (2020-2022)
sensitivities at $1\sigma$ C.L., respectively. The blue curve is
the GWs background generated by all binary black hole coalescence
without excluding potentially resolvable binaries.  }
\label{fig01}
\end{figure}

\begin{figure}[htbp]
\includegraphics[scale=2,width=0.55\textwidth]{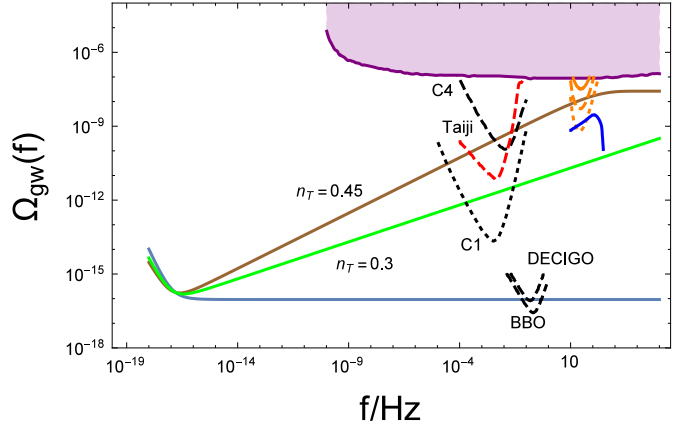}
\caption{The green and the brown lines are the stochastic GWs
backgrounds from inflation with $n_T=0.3$ in (\ref{para1}) and
$n_T=0.45$ in (\ref{para2}), respectively. Both C1 and C4-lines
are eLISA's representative configurations given in
\cite{Caprini:2015zlo}. The sensitivity curves of DECIGO and BBO are given in \cite{Kuroyanagi:2010mm}. The red dashed curve is Taiji's sensitivity curve,
see, e.g., \cite{Gao:2016tzv} for a preliminary report. Fig.\ref{fig01}
is actually the amplification of image at the frequency band
10-400 Hz in this figure.} \label{fig02}
\end{figure}

We plot the stochastic background of our GWs in Fig.\ref{fig01}.
It is obvious that a blue-tilt primordial GWs with $n_T\gtrsim
0.4$ is able to contribute a large stochastic GWs background within the
windows of Advanced LIGO/Virgo, which may be greater than the contribution from the
incoherent superposition of all binary black hole coalescence.
%Actually, this astrophysical spectrum has a different power-law
%dependence $\Omega_{gw}\sim k^{2/3}$.
$n_T\gtrsim 0.4$ requires $p\gtrsim 2/3$ in (\ref{cTn}), which
suggests that the diminishment of $c_T$ in units of Hubble time is
not too fast. It is also interesting to notice that if such a GWs
background could be detected by Advanced LIGO/Virgo in upcoming
observing runs, it will also be able to be detected by the
space-based interferometers at a lower frequency band, such as eLISA,
and China's Taiji program in space, see Fig.\ref{fig02}, as well
as the PTA, e.g.\cite{Lasky:2015lej}\cite{Liu:2015psa}.

%\begin{figure}[htbp]
%\includegraphics[scale=2,width=0.55\textwidth]{nt_0.4.eps}
%\caption{The CMB B-mode spectrum for the primordial GWs with
%$n_T=-r/8$ in (\ref{para1}) and $n_T=0.4$ in (\ref{para2}),
%respectively. } \label{fig02}
%\end{figure}

\section{Disformal dual to superinflation }\label{app-disformal}

\begin{figure}[htbp]
\includegraphics[scale=2,width=0.55\textwidth]{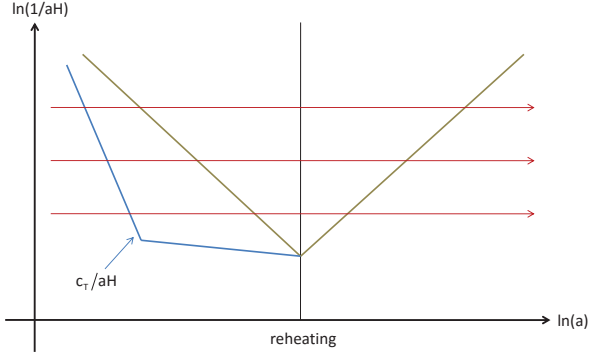}
\caption[Caption for LOF]{This sketch illustrates the evolutions of the primordial
perturbations during inflation in our scenario. The brown line is
$\sim 1/aH$. The blue line is $\sim c_T/aH$, which is the sound
horizon of GWs.
We assume that $c_T$ decrease to some value less than unit and begin to increase later, so that it could return to unity and both horizons coincides before or near the end of inflation. }\label{fig03}
\end{figure}

%\footnote{In Fig. \ref{fig02}, $\Omega_{gw}$ may begin to decrease at higher frequency (e.g., $f>10^4$ Hz) due to the increasing of $c_T$ near the end of inflation. We haven't plot it since we are mainly interested in the blue spectrum of GWs here.}

%The perturbation mode with $k\ll a H_{Per}$ will freeze,
%while it will evolve inside $1/(aH_{Per})$, where
%$1/(aH_{Per})$ is the comoving sound horizon of the
%perturbations. In inflation scenario, the spectrum of GWs
%generally has similar shape to that of the scalar perturbation,
%since both $1/(aH_{Per})$ almost coincide with $1/(aH)$. Here,
%since the comoving sound horizon of GWs is $c_T/(aH)$, and its
%evolution is completely different from $1/(aH)$, the spectrum of
%GWs shows itself blue-tilt, see Fig.\ref{fig03}.
%
%The superinflation is the inflation with $\epsilon=-{\dot
%H}/H^2<0$, i.e. ${\dot H}>0$, which breaks the NEC. The tilt of
%$c_T/(aH)$-line in Fig.\ref{fig03} is same with that of the superinflation
%with $c_T=1$, see Fig.1 in \cite{Piao:2006jz}. Here, we will
%clarify the corresponding relation.

The superinflation is the inflation with $\epsilon=-{\dot
H}/H^2<0$, i.e. ${\dot H}>0$, which breaks the NEC. The model we proposed in Sec. \ref{model}, i.e., inflation with
a diminishing $c_T=(-H_{inf}\tau)^p$, is actually disformally dual to superinflation. This can be inferred from the evolution of the GWs sound horizon.

The perturbation mode outside the comoving sound horizon $1/(aH_{Per})$ of the
perturbations\footnote{Here, $H_{Per}$ is defined as $H_{Per}={(z_T''/z_T)^{1/2}\over a c_T}$, where $z_T$ is given by Eq.(\ref{zt}). For $c_T\sim (-\tau)^p$ and $a\sim(-\tau)^{-1}$, we have
\be {1\over aH_{Per}}=\sqrt{1\over 2+3p+p^2}{c_T\over aH}\sim{c_T\over aH}.
\ee }
(i.e., $k\ll a H_{Per}$) will freeze,
while it will evolve inside $1/(aH_{Per})$. In inflation scenario, the spectrum of GWs
generally has similar shape to that of the scalar perturbation,
since both GWs and scalar perturbations have a comoving sound horizon $1/(aH_{Per})$ almost coincide with $1/(aH)$. Here,
since the comoving sound horizon of GWs is $c_T/(aH)$, and its
evolution is completely different from $1/(aH)$, the spectrum of
GWs shows itself blue-tilt, see Fig.\ref{fig03}.
However, the tilt of
$c_T/(aH)$-line in Fig.\ref{fig03} is the same as that of the superinflation
with $c_T=1$, see Fig.1 in \cite{Piao:2006jz}. This indicates the physical processes of horizon crossing of GWs modes are same in these two scenarios, thus will generate the same power spectra. In fact, these two scenarios can be connected by a disformal transformation. Bellow, we give the strict proof.

We make a disformal redefinition of the metric
\cite{Creminelli:2014wna} \be g_{\mu\nu}\rightarrow
c_T^{-1}\left[g_{\mu\nu}+{(1-c_T^2)}n_{\mu}n_{\nu}\right]\,.
\label{gmunu1} \ee with \be \tilde{t}\equiv\int c_T^{1/2}dt,\qquad
\tilde{a}(\tilde{t})\equiv c_T^{-1/2}a(t), \label{tildea}\ee which
makes (\ref{paction}) become \be S^{(2)}=\int d\tilde{\tau} d^3x
{M_p^2\tilde{a}^2\over 8}\lf[\lf(\frac{d\gamma_{ij}
}{d\tilde{\tau}}\rt)^{\,2}-(\vec{\nabla}\gamma_{ij})^2\rt]\,\label{newaction}
\ee with ${\tilde c_T}=1$.

Here, with $d\tilde{\tau}={d\tilde{t}/\tilde{a}}$, which implies
\be \tilde{\tau}=\int^{\tau}(-H_{inf}\tau)^p
d\tau=-(H_{inf})^p{(-\tau)^{p+1}\over p+1}, \ee we have \be
\tilde{a}=c_T^{-1/2}a \sim(-\tilde{\tau})^{-{2+p\over 2(1+p)}}.
\ee \be \tilde{H} =\frac{d\tilde{a}/d\tilde{t} }{\tilde{a}
}=c_T^{-1/2}\lf(H_{inf}-\frac{dc_T/dt}{2c_T}\rt)\sim
(-\tilde{\tau})^{-{p\over2(1+p)}}. \label{tildeH}\ee Thus the
value of $\tilde H$ is gradually increasing. This suggests that after the disformal transformation
the background is actually the superinflation with ${\tilde
\epsilon}=-p/(2+p)$, which satisfies $-1\lesssim
\tilde{\epsilon} <0$. The scenario with $\tilde{\epsilon}\ll -1$
is the slow expansion, which was implemented in
\cite{Piao:2003ty}.

The equation of motion for $u(\tilde{\tau},k)$ is \be
\frac{d^2u}{d\tilde{\tau}^2
}+\left(k^2-\frac{d^2\tilde{z}_T/d\tilde{\tau}^2}{\tilde{z}_T}
\right)u=0,\label{eom2} \ee where ${u}(\tilde{\tau},k)=
\gamma_{\lambda}(\tilde{\tau},k) {\tilde{z}_T}$ and $\tilde{z}_T=
{\tilde{a}M_p/ 2}$. The initial state is still the Bunch-Davies
vacuum $u\sim\frac{1}{\sqrt{2k} }e^{-ik\tilde{\tau}}$. The
solution is \be u_k(\tilde{\tau})={\sqrt{\pi}\over2\sqrt{k}}
\sqrt{-k\tilde{\tau}}H^{(1)}_{\tilde{\nu}}(-k\tilde{\tau}), \ee
%where $\tilde{\nu}=1+{1\over2(1+p)}$ and
%\be
%H^{(1)}_{\tilde{\nu}}(-k\tilde{\tau})\overset{-k\tilde{\tau}\rightarrow0}
%\approx-i\lf({2\over -k\tilde{\tau}}\rt)^{1+{1\over
%2(1+p)}}\cdot{\Gamma\lf(1+{1\over 2(1+p)}\rt)\over \pi}\,\,.
%\ee
where
\be
H^{(1)}_{\tilde{\nu}}(-k\tilde{\tau})\overset{-k\tilde{\tau}\rightarrow0}
\approx-i\lf({2\over -k\tilde{\tau}}\rt)^{\tilde{\nu}}\cdot{\Gamma\lf(\tilde{\nu}\rt)\over \pi}\,\,,
\ee
and $\tilde{\nu}=1+{1\over2(1+p)}$.
Thus the power spectrum is \ba P_T &= &
\frac{k^3}{2\pi^2}\sum_{\lambda=+,\times} \lf|\gamma_{\lambda}
\rt|^2 \nn\\ &=& {4k^3\over \pi^2M_p^2\tilde{a}^2}\cdot{\pi\over
4k}(-k\tilde{\tau})
{2^{2+{1\over1+p}}\over (-k\tilde{\tau})^{2+{1\over 1+p}}}\cdot{1\over4(1+p)^2}{\Gamma^2\lf({1\over2(1+p)}\rt)\over \pi^2}\nn\\
&=&{2\tilde{H}^2\over \pi^2 M_p^2}\cdot {2^{1+{1\over1+p}}\over\pi
(2+p)^2}
\Gamma^2\lf({1\over2(1+p)}\rt)(-k\tilde{\tau})^{p\over1+p}\label{Ptilde}\\
&=&{c_Tk^2\over\pi^3M_p^2}\cdot {(-\tau)^2 H_{inf}^2
2^{1\over1+p}\over(1+p)^2}\Gamma^2\lf({1\over2(1+p)}\rt)
\lf(k\cdot{(-\tau)^{1+p}\over1+p}H_{inf}^p\rt)^{-{2+p\over1+p}}\nn\\
&=&{2H_{inf}^2\over\pi^2M_p^2c_T}\cdot{2^{-p\over1+p}\over\pi}
\Gamma^2\lf({1\over2(1+p)}\rt)(-ky)^{p\over1+p}. \ea This result
is completely the same as Eq.(\ref{nT}).

When $p\ll1$, we have  \be {2^{1+{1\over1+p}}\over\pi(2+p)^2}
\Gamma^2\lf({1\over2(1+p)}\rt)\approx1+0.27p+{\cal O}(p^2) \ee in
Eq.(\ref{Ptilde}) and $\tilde{\epsilon
}=-\frac{d\tilde{H}/d\tilde{t} }{\tilde{H}^2}\ll 1$. Thus with
(\ref{Ptilde}), we have \be P_T=2{\tilde H}^2/\pi^2 M_P^2,
\label{PT3}\ee i.e. Creminelli \textit{et.al}'s result
\cite{Creminelli:2014wna}.

Actually, it is well known that the spectrum of GWs, as well as
scalar perturbation, is independent of the disformal redefinition
(\ref{gmunu1}) of the metric
\cite{Creminelli:2014wna}\cite{Minamitsuji:2014waa}. An intuition
argument for it is the comoving horizon of scalar perturbation
\be {{\tilde c}_s \over \tilde{a}\tilde{H}}= {1\over
c_T\tilde{a} \tilde{H}}\sim(-\tilde{\tau})^{1\over1+p}\sim {1\over
a H_{inf}}
\ee
%{\red
%\be c_T\tilde{H}=\lf(1+{p\over2}\rt)c_T^{1/2}H_{inf}
%\ee}
i.e., the relation between the comoving wave number $k$ and the
comoving sound horizon is not altered, where ${\tilde c}_s=1/c_T$
\cite{Creminelli:2014wna}.

Conventionally, the superinflation breaks the NEC. Implementing the superinflation without the ghost instability is
still a significant issue,
e.g.\cite{Wang:2014kqa}\cite{Cai:2014uka}. Here, we actually
suggest such a superinflation scenario. It might be just a
slow-roll inflation living in a disformal metric with $c_T$
gradually diminishing, however, if we see it with $c_T=1$, what we
will feel is the superinflation. The violation of NEC in modified gravity does not necessarily mean ghost instability. Because the quadratic actions (\ref{paction}) and (\ref{newaction}) for the tensor (as well as those for scalar) are canonical, there is no ghost
instability in both frames.

\section{Cutoff of blue spectrum}\label{app-cutoff}

To avoid $P_T\sim 1$ at high frequency, we have to require that
the diminishment of $c_T$ stop at a certain time $\tau_c$.
Additionally, we assume that $c_T(t)$ will return to unity before the end of inflation, as in Sec. \ref{twoB}.

\begin{figure}[htbp]
\includegraphics[scale=2,width=0.55\textwidth]{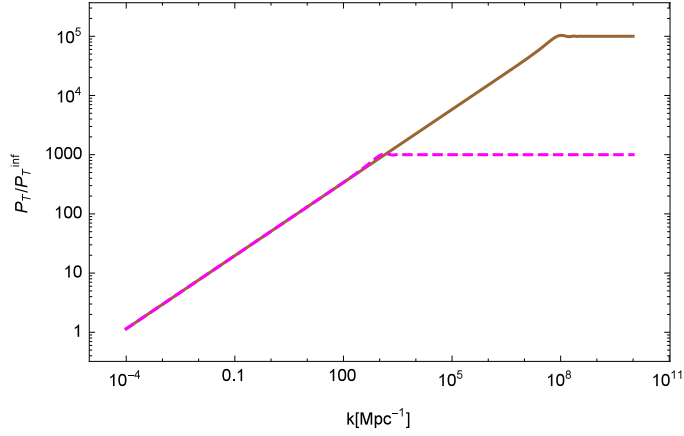}
\caption{$P_T/P_{T}^{inf}=f(p,y_c,k)$. The parameters of the
magenta dashed and brown solid curves are
$c_{Tc}=10^{-3}$ and $10^{-5}$, respectively, while we set $p=0.7$.}\label{fig04}
\end{figure}

We assume that \ba c_T&=&(-H_{inf}\tau)^p\,\,\,\,\mathrm{for}\,\,\,\tau<\tau_c,\nn\\
c_T&=&c_{Tc}\qquad\quad\,\,\mathrm{for}\,\,\,\tau>\tau_c. \ea We set
$dy=c_Td\tau$. The solution of (\ref{eom4}) is \be
u_2(y)=\sqrt{-ky}\lf[C_1(k) H_{3/2}^{(1)}(-ky)+C_2(k)
H_{3/2}^{(2)}(-ky)\rt] \ee for $y>y_c$, and is $u_1(y)$ for
$y<y_c$, which is actually (\ref{cq002}), where $\nu=1+{1\over
2(1+p)}$, $y_c=c_{Tc}\tau_c$. When $-ky\ll1$, \be
u_2\approx{\sqrt{2}\over-ky\sqrt{\pi}}|C_1-C_2|. \ee Thus the
spectrum of primordial GWs is
\ba P_T&=&{4k^3\over \pi^2 M_P^2}
{c_T|u|^2\over a^2}= {2H_{inf}^2\over\pi^2M_p^2} f(p,y_c,k),
\label{PT4}
\ea
where
\ba f(p,y_c,k)&=&{4k\over\pi
c_{Tc}}|C_1-C_2|^2,
\ea
and
\ba
C_1&=&-{i\pi^{3/2}\over16\sqrt{k}}\Big[-2ky_cH^{(1)}_{\nu-1}(-ky_c)
H^{(2)}_{3/2}(-ky_c)\nn\\
&\,&\qquad\qquad+H^{(1)}_{\nu}(-ky_c)
\lf(2ky_cH^{(2)}_{1/2}(-ky_c)+(3-2\nu)H^{(2)}_{3/2}(-ky_c)\rt)\Big],
\ea
\ba
C_2&=&{i\pi^{3/2}\over16\sqrt{k}}\Big[2ky_cH^{(1)}_{\nu}(-ky_c)
H^{(1)}_{1/2}(-ky_c)\nn\\
&\,&\qquad\qquad+H^{(1)}_{3/2}(-ky_c)
\lf(-2ky_cH^{(1)}_{\nu-1}(-ky_c)+(3-2\nu)H^{(1)}_{\nu}(-ky_c)\rt)\Big]
\ea are set by the continuities of $u(y)$ and ${du/ dy}$ at
$\tau_c$. We plot (\ref{PT4}) in Fig.\ref{fig04}, and see that,
although $P_T$ has a blue tilt, it is flat at a high frequency. We
analytically calculate it as follows.

When $-ky_c\ll1$,
\ba
C_1&=&-2^{-{4+5p\over2(1+P)}}e^{iky_c}
{1\over\sqrt{k}}(-ky_c)^{-{6+5p\over2(1+p)}}
{\Gamma\lf(3+2p\over2(1+p)\rt)\over1+p}
\nn\\
&\,&\cdot\lf[ip+pky_c+2ip(1+p)(ky_c)^2+2(1+p)^2(ky_c)^3\rt], \ea
\ba
C_2&=&2^{-{4+5p\over2(1+P)}}e^{-iky_c}{1\over\sqrt{k}}
(-ky_c)^{-{6+5p\over2(1+p)}}{\Gamma\lf(3+2p\over2(1+p)\rt)\over1+p}
\nn\\
&\,&\cdot\lf[-ip+pky_c-2ip(1+p)(ky_c)^2+2(1+p)^2(ky_c)^3\rt]. \ea
We have
\ba f(p,y_c,k)&=&{4k\over\pi c_{Tc}}|C_1-C_2|^2\nn\\
&\simeq &{2^{-{p\over1+p}}\over9(1+p)^2\pi
c_{Tc}}\Gamma^2\lf({3+2p\over2(1+p)}\rt)
(6+5p)^2(-ky_c)^{{p\over1+p}}. \ea Thus the tilt $n_T={p\over
1+p}$, which is the same as (\ref{nT}).

When $-ky_c\gg1$, \ba C_1=e^{{i\pi\over4}-{i\nu\over2}\pi}\cdot
{\sqrt{\pi}\over
8k^{5/2}y_c^2}\lf[i(2\nu-3)+(2\nu-5)ky_c+4i(ky_c)^2\rt], \ea \ba
C_2=e^{{i\pi\over4}-{i\over2}(\pi\nu+4ky_c)}\cdot {\sqrt{\pi}\over
8k^{5/2}y_c^2}\lf[i(2\nu-3)+(1-2\nu)ky_c\rt]. \ea We have \ba
f(p,y_c,k) &\approx&{1\over c_{Tc}}. \ea Thus the spectrum is
flat.

From the above result, we can infer that if $c_T$ slowly diminishes to a value less than unity during inflation and then increases back to unity before the end of inflation, $\Omega_{gw}$ could be strongly boosted at the frequency  band of Advanced LIGO/Virgo.

\section{Discussion}

In the inflation scenario, obtaining a blue GWs spectrum
($n_T>0.1$) without the ghost instability while reserving a
scale-invariant scalar spectrum with a slightly red tilt is still a
challenge. We find that if the propagating speed of GWs gradually
diminishes during inflation, the power spectrum of primordial GWs
will be strongly blue, while that of the scalar perturbation may be
unaffected.

It is well known that the blue-tilt GWs is the hallmark of 
superinflation \cite{Piao:2004tq}\cite{Baldi:2005gk}. It may be
implemented without ghost in G-inflation \cite{KYY},  but it is difficult,  however, to simultaneously give it a slightly red-tilt
scalar spectrum \cite{Wang:2014kqa}, see also \cite{Cai:2014uka}.
Our scenario is actually a disformal dual to the superinflation,
%The
%background in our mechanism equals to the superinflation with
%$c_T=1$,
see Sec. \ref{app-disformal}. In this duality, our background is actually a
slow-roll inflation living in a disformal metric with $c_T$
gradually diminishing. However, if we see it with $c_T=1$, what we will
feel is the superinflation, but there is no ghost instability.
Thus our work might offer a far-sighted perspective on
superinflation.

The blue tilt obtained is $0<n_T\lesssim 1$, which may
significantly boost the stochastic GWs background at the frequency
band of Advanced LIGO/Virgo, as well as the space-based detectors.
This indicates that the primordial GWs recording the origin of the
universe may be potentially measurable by the corresponding experiments.

To conclude, if a stochastic
background of GWs is detected by Advanced LIGO/Virgo  in the upcoming observing runs, it also possibly
comes from the primordial inflation, and encodes the physics
beyond GR at inflation scale.

%The physical origin of $c_T=c_T(t)$ is significant. It could be
%implemented in modified gravity, which is interesting for study.
%In addition, it is also interesting to link it to the string and
%supergravity theory. Therefore, such a stochastic GWs background
%actually encodes the physics beyond GR at inflation scale.

\textbf{Acknowledgments}

This work is supported by NSFC, No. 11222546, 11575188, and the
Strategic Priority Research Program of Chinese Academy of
Sciences, No. XDA04000000. We thank Cong-Feng Qiao and Yun-Kau Lau
for suggesting that we use the sensitivity curves of China's Taiji
program in space, which will appear in a full work report.

\appendix

\section{Scalar perturbation}\label{scalar}

%We work in the ADM formalism, the metric is \be
%ds^2=-N^2dt^2+h_{ij}(N^idt+dx^i)(N^jdt+dx^j)\,, \ee

We work with the ADM metric
\begin{eqnarray}
 g_{\mu\nu}=\left(\begin{array}{cc}N_kN^k-N^2&N_j\\N_i& h_{ij}\end{array}\right),~~
 g^{\mu\nu}=\left(\begin{array}{cc}-N^{-2}&\frac{N^j}{N^2}\\ \frac{N^i}{N^2}& h^{ij}-\frac{N^iN^j}{N^2}\end{array}\right)\,,
\end{eqnarray}
where $h_{ij}=a^2e^{2\zeta}(e^{\gamma})_{ij}$, and
$\gamma_{ii}=0=\partial_i\gamma_{ij}$. Generally, $N=1+\alpha$ and
$N_i=\partial_i\beta$ are set for the scalar perturbations . It is
convenient to define the normal vector of 3-dimensional
hypersurface $n_\mu=n_0{dt/dx^\mu}=(n_0,0,0,0)$ and
$n^{\mu}=g^{\mu\nu}n_\nu$. Using the normalization $n_\mu
n^\mu=-1$, one has $n_0=-N$, which suggests $n_\mu=(-N,0,0,0),
n^\mu=(\frac{1}{N},\frac{N^i}{N})$, and the 3-dimensional induced
metric, orthogonal to the normal vector, i.e., $H_{\mu\nu}n^\nu=0$,
can be defined to be $H_{\mu\nu}=g_{\mu\nu}+n_\mu n_\nu$,
\begin{eqnarray}
 H_{\mu\nu}=\left(\begin{array}{cc}N_kN^k&N_j\\N_i& h_{ij}\end{array}\right),~~
 H^{\mu\nu}=\left(\begin{array}{cc}0&0\\ 0& h^{ij} \end{array}\right).
\end{eqnarray}
%The induced 3-dimensional metric on the supersurface is
%$H_{\mu\nu}=g_{\mu\nu}+n_{\mu}n_{\nu}$, where $n_{\mu}$ is the
%unit normal vector of the 3-dimensional hypersurface,
%$n_{\mu}n^{\nu}=-1$.
The covariant derivative associated with $H_{\mu\nu}$ is
$D_{\mu}$, which is applied to define the extrinsic curvature
$K_{\mu\nu}$: 
\ba K_{\mu\nu}={1\over
2N}(\dot{H}_{\mu\nu}-D_{\mu}N_{\nu}-D_{\nu}N_{\mu}). \ea We have
\ba & & \delta K_{\mu\nu}\delta K^{\mu\nu}-(\delta K)^2\nn\\
&=&{1\over (1+\alpha)^2}\Big\{-6(\dot{\zeta}-\alpha H)^2
+4a^{-2}e^{-2\zeta}(\dot{\zeta}-\alpha
H)(\partial_i\partial_i\beta
+\partial_i\beta\partial_i\zeta)\nn\\
%%%
&\,& \qquad\qquad\,+a^{-4}e^{-4\zeta}\Big[
(\partial_i\partial_j\beta-\partial_i\beta\partial_j\zeta-\partial_j\beta\partial_i\zeta)^2
-2(\partial_i\beta\partial_i\zeta)^2-(\partial_i\partial_i\beta)^2
\Big] \Big\}\,, \ea
where $\delta K_{\mu\nu}=K_{\mu\nu}-H_{\mu\nu}H$.

Thus the quadratic action of scalar perturbation for
(\ref{action1}) and (\ref{action}) is \ba
 &\,&S^{(2)}_\zeta=\int dx^4 M_p^2  \Big\{ a^3H^2 \alpha ^2 \epsilon
 - 27a^3H^2 \zeta ^2
 + 9a^3H^2 \epsilon  \zeta ^2
 %+ {\blue 2 a H \zeta  \partial^2\beta}
  - 18a^3H \zeta  \dot\zeta\nn\\
 %+{\blue 2 a H \partial_i\beta  \partial_i\zeta}
 &\,&\qquad\qquad\qquad\qquad \,\, +a\left(\partial \zeta
 \right)^2
 %{\green - a \left(\partial \zeta \right)^2}
 - 2 a\alpha  \partial_i\partial_i\zeta
 %{\green - 2 a \zeta  \partial^2\zeta}
 %%%%%%%%%
 -{1\over c_T^2} \Big[ 3a^3H^2 \alpha ^2
 -  6a^3H \alpha  \dot\zeta
 +  3 a^3\dot\zeta {}^2\nn\\
&\,&\qquad\qquad\qquad\qquad \,\,
 -  2 a\partial_i\partial_i\beta ( \dot\zeta
 -H \alpha )
 %{\magenta -  {\left(\partial_i\partial_j\beta \right)^2 \over 2a} }
% {\magenta + { (\partial^2\beta)^2     \over 2a} }
   \Big]
 %%%%%%%%
 \Big\}\,.
\label{scalar1}\ea The constraints can be solved as
\ba \alpha&=&{\dot{\zeta}\over H}\,,\\
%%%%%%%%%%%%%%%%%
\partial_i\partial_i\beta&=&{c_T^2\over H}(a^2H\epsilon\dot{\zeta}-\partial_i\partial_i\zeta)\,.
\ea 
Inserting them into (\ref{scalar1}), \ba S^{(2)}_\zeta=\int dx^4
{M_p^2} a^3\epsilon \lf[\dot{\zeta}^2-{(\partial\zeta)^2\over
a^2}\rt]\, \ea
is obtained. Therefore, the scalar perturbation is
not affected by the operator $\delta K_{\mu\nu}\delta
K^{\mu\nu}-(\delta K)^2$ at quadratic order.

 \end{document}